\documentclass{article}
\usepackage{spconf,amsmath,graphicx}
\usepackage{url}

\newcommand{\etal}{\textit{et al.\,}}
\newcommand{\eg}{\textit{e.g.}}
\newcommand{\ie}{\textit{i.e.}}

\usepackage{xcolor}
\usepackage{bbding}
\usepackage{pifont}

\usepackage{cleveref}
\crefname{appendix}{App.\negthinspace\,}{App.\negthinspace\,}
\crefname{chapter}{Chap.\negthinspace\,}{Chap.\negthinspace\,}
\crefname{equation}{Eq.\negthinspace\,}{Eq.\negthinspace\,}
\crefname{algorithm}{Alg.\negthinspace\,}{Alg.\negthinspace\,}
\crefname{section}{Sec.\negthinspace\,}{Sec.\negthinspace\,}
\crefname{subsection}{Sec.\negthinspace\,}{Sec.\negthinspace\,}
\crefname{subsubsection}{Sec.\negthinspace\,}{Sec.\negthinspace\,}
\crefname{figure}{Fig.\negthinspace\,}{Fig.\negthinspace\,}
\crefname{table}{Tab.\negthinspace\,}{Tab.\negthinspace\,}
\crefname{subfigure}{Fig.\negthinspace\,}{Fig.\negthinspace\,}
\crefname{subsubfigure}{Fig.\negthinspace\,}{Fig.\negthinspace\,}
\crefname{lstlisting}{Lst.\negthinspace\,}{Lst.\negthinspace\,}

\title{Semi-Automatic Generation of Tight Binary Masks and Non-Convex Isosurfaces for Quantitative Analysis of 3D Biological Samples}
\name{Sourabh Bhide$^{1,2}$, Ralf Mikut$^{3}$, Maria Leptin$^{4}$, Johannes Stegmaier$^{5}$ \thanks{We thank Alexandre Cunha (California Institute of Technology, Pasadena, CA, USA) and Thiago Vallin-Spina (Brazilian Synchrotron Light Laboratory (LNLS) and Brazilian Center for Research in Energy and Materials (CNPEM), Campinas, Brazil) for providing access to SEGMENT3D.}}
\address{
\footnotesize $^{1}$Director's Research Unit, European Molecular Biology Laboratory, Heidelberg, Germany\\
\footnotesize $^{2}$Collaboration for joint PhD degree between EMBL and Heidelberg University, Faculty of Biosciences\\
\footnotesize $^{3}$Institute for Automation and Applied Informatics, Karlsruhe Institute of Technology, Karlsruhe, Germany\\
\footnotesize $^{4}$European Molecular Biology Organization, Heidelberg, Germany\\
\footnotesize $^{5}$Institute of Imaging and Computer Vision, RWTH Aachen University, Aachen, Germany\\
}
\begin{document}
%
\maketitle
\begin{abstract}
Current \emph{in vivo} microscopy allows us detailed spatiotemporal imaging (3D+t) of complete organisms and offers insights into their development on the cellular level. Even though the imaging speed and quality is steadily improving, fully-automated segmentation and analysis methods are often not accurate enough. This is particularly true while imaging large samples ($100 \operatorname{\mu m} - 1 \operatorname{mm}$) and deep inside the specimen. \emph{Drosophila} embryogenesis, widely used as a developmental paradigm, presents an example for such a challenge, especially where cell outlines need to imaged – a general challenge in other systems as well. To deal with the current bottleneck in analyzing quantitatively the 3D+t light-sheet microscopy images of \emph{Drosophila} embryos, we developed a collection of semi-automatic open-source tools. The presented methods include a semi-automatic masking procedure, automatic projection of non-convex 3D isosurfaces to 2D representations as well as cell segmentation and tracking.
\end{abstract}
\begin{keywords}
3D+t Image Analysis, Segmentation, Tracking, Visualization, Developmental Biology, GUI
\end{keywords}
\section{Introduction}
\label{sec:intro}
Analysis of the 3D organisation of multicellular organisms and organ systems is the basis for understanding their morphogenesis and function. Deciphering how the constituent individual cells or groups of cells coordinate their shape and number is necessary to describe how the 3D architecture of a living organism is established. Advancement in high temporal and spatial \emph{in vivo} imaging microscopes has facilitated the acquisition of the data needed to achieve a complete description. However, to understand events at multicellular scales, one needs to visualize the entire sample simultaneously. Although several 3D segmentation methods were successfully applied for studying biological samples in a quantitative manner \cite{Khan14, Amat14, Stegmaier16g, Wolny2020}, these methods are often limited by the quality of the acquired images. Image quality deteriorates rapidly at deeper imaging planes ($30 - 50 \operatorname{\mu m}$) due to diffraction caused by cellular and extracellular material within the sample \cite{Helmchen05, Mavrakis08, Willis2016}. Thus, automatic segmentation in such images is still error-prone and especially in the deeper layers, and in the case of rapid movements, low signal intensity or very fine or small structures, segmentation becomes ambiguous and causes commonly used methods like the watershed to erroneously flood into adjoining cells or the background \cite{Mosaliganti12, Eschweiler19}. \newline
\begin{figure*}[htb]
    \includegraphics[width=\textwidth]{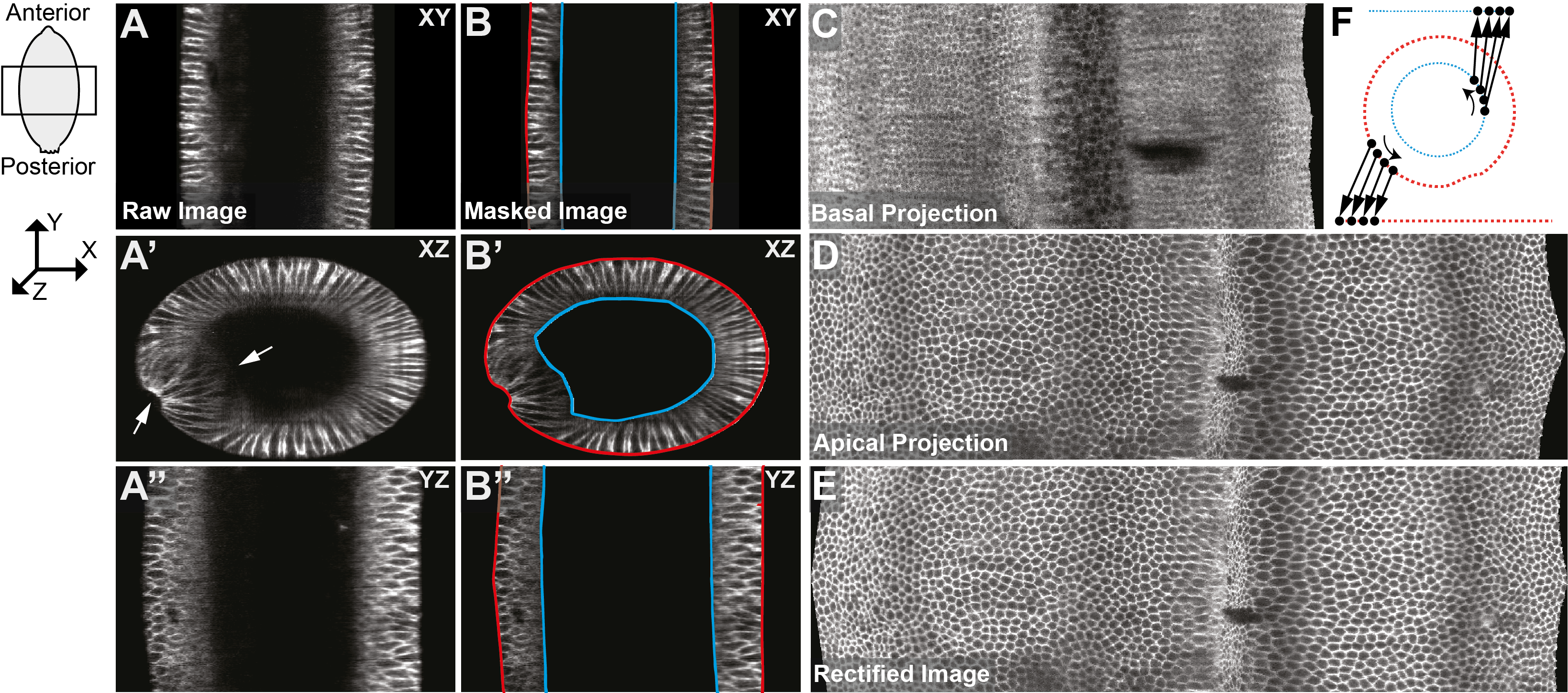}
    \vspace{-0.7cm}
    \caption{(A) Cross-sections of 3D light-sheet microscopy images of a \emph{Drosophila} embryo along the XY-, XZ- and YZ-plane. Arrows indicate non-convex tissue at the ventral furrow. (B) Semi-automatically identified apical (red) and basal (cyan) surfaces of the raw image in (A). (C, D) Apical and basal peels extracted based on the mask of step (B). (E) Rectified apical peel with the ventral furrow centered and straightened. (F) Schematic illustration of the 3D surface to 2D peel projection. }
    \label{fig:Figure1}
\end{figure*}
One way of simplifying the quantification of 3D images is to analyze 2D sections. Orthogonal sections have been successfully used to extract cell and tissue shape changes. However, a particular problem is posed by the fact that biological specimens usually have curved surfaces. For such structures, the orthogonal sections are often inaccurate and depend on the selection of the orthogonal plane. One way of dealing with this problem has been to use cartographic projections to represent 3D isosurfaces on 2D planes \cite{Krzic12, Schmid13, Heemskerk15, Kobitski15}, in the same way as the spherical earth is represented on a 2D map. Similar approaches are also found in other disciplines, \eg, in image-based quality control of critical infrastructure \cite{Lai2016}. These projections, however, have a number of constraints. They work properly only when supplied with a highly accurate mask of the foreground region from which to extract the isosurfaces. Although the identification of the background may appear trivial, the automatic extraction of binary masks that accurately separate
the background from the foreground regions is a challenging and largely unsolved problem. Most importantly, existing methods make the assumption that specimens have simple parametric shapes like ellipsoids \cite{Krzic12} or spheres \cite{Schmid13}, or depend on a sufficiently accurate segmentation being provided from a previous processing step \cite{Kobitski15, Heemskerk15, Stegmaier18}. Moreover, cartographic projections inevitably distort the projected data at the poles. While this can be partly compensated by performing quantification in the original space \cite{Heemskerk15}, the visual appearance is still affected. In addition to enabling the extraction of accurate, potentially non-convex, isosurfaces, a precise specification of the region of interest to be analyzed can also help to improve the results of both manual and automatic segmentation and tracking methods in noisy 3D+t image data. To overcome some of these problems, we implemented a collection of tools that enable an extensive semi-automatic analysis of the early \emph{Drosophila} embryo. The protocol includes tools for (1) semi-automatic 3D mask generation, (2) projection of non-convex 3D isosurfaces to 2D peels, (3) a peel rectification module and (4) methods to segment, track and quantify.

\section{Generation of Tight Binary Masks and Non-Convex Isosurfaces}
\subsection{Semi-Automatic Mask Generation}
\label{sec:MaskGeneration}
Estimating tight binary masks in regions with low signal to noise ratio at the interface between basal membranes of the cells and the auto-fluorescing interior of the embryo is currently impossible with automatic methods. Our first module attempts to solve this issue by using a semi-automatic approach for an accurate delineation of the transition region at the apical and the basal surface (\cref{fig:Figure1}A). We oriented the image stacks consistently, such that the left-right axes aligned with the x-axis, the anteroposterior axis aligned with the y-axis and the dorsoventral axis aligned with the z-axis of the image space. We focused on cylindrical crops from the center part of the ellipsoidal embryo (\cref{fig:Figure1}), where the cell mass of interest resembles a hollow cylinder. To facilitate the annotation procedure and to avoid costly manual labeling of each of the slices separately, the GUI extracts the first and the last slice of a cross-section through the embryo. The user is then requested to draw the outer and the inner boundary in the first and the last slice, respectively, using either polylines or free hand drawing (\cref{fig:Figure1}B). Based on these masks for the outer slices of the stack, we interpolate the masks of intermediate slices using linear interpolation. In cases where the embryo significantly deviated from this linearity assumption (\ie, where a linear interpolation between the masks would not be possible), the script allows to additionally annotate a desired amount of intermediate slices and we sequentially perform a piecewise linear interpolation between successive masks to obtain the final masked image crop. On the one hand, the masks are used to select the image regions of interest and on the other hand, we add the surface pixels of the mask as a safety margin to the raw images. This safety margin helps to improve the segmentation step, by preventing watershed-based flooding of single cells from leaking into background areas and vice versa.

\subsection{Surface Peel Extraction and Projection}

For visualization purposes and to measure cell shape properties on the apical and basal surface of the embryo, respectively, we developed an additional peel extraction module that unwraps the surfaces of the embryo to a 2D projection. Using the manually created mask (central part of the embryo spanning the entire dorsoventral extent), we create a Euclidean distance map (EDM) of the mask image (\ie, the intensity of each foreground voxel is assigned the Euclidean distance to the closest background pixel). We extract the apical and the basal surface by thresholding the EDM using a threshold range of $t = 5 \pm 0.5$ px, such that only pixels with a Euclidean distance to the background within this range are selected. We then identify the segment corresponding to the apical surface by selecting the connected component with the largest volume. The second largest connected component is used as the basal surface. To unwrap the embryo and to project the 3D surface to a 2D image, we unwrap the crop in a slice-by-slice fashion on cross-section slices (x-axis: left-right, y-axis: dorsoventral). Starting at an arbitrary point, the surface of the current shell is traversed in a clockwise fashion and we successively fill up one row of the unwrapped 2D image (\cref{fig:Figure1}F). Once one full revolution of the current slice is traversed, we continue the extraction at the next slice and use the spatially nearest neighbor of the previous starting point as the starting point for the new slice (\cref{fig:Figure1}C, D). As we do not limit the extraction to a particular geometric shape, we are able to also unwrap non-convex shapes like the ones observed during ventral furrow formation (white arrows in \cref{fig:Figure1}A). This procedure is analogous to stretching a folded rubber band to a planar representation. To avoid holes in the 2D projections that are caused by the discretization of the performed steps during the boundary tracing, we linearly interpolate missing values using the respective spatially closest neighbor pixels. Anisotropic input images are upsampled in the axial direction as a preprocessing step before the peel extraction, to preserve physically correct cell shapes. By explicitly tracing the boundary and by treating vertical and horizontal steps \emph{vs.} diagonal steps differently, we can measure length and area in the 2D projections even for cross sections of non-convex tissues without having to deal with issues like distortions at poles that are frequently observed for cartographic projections. To compensate for the more or less arbitrary orientation of the embryo in the 3D images, we implemented a module that can be used to align the ventral furrow (or other linear structures of interest) in the center of the projected image, just like centering text in word processing software. This step is performed semi-automatically by letting the user add click points along the ventral furrow. The manually drawn line is used to automatically center and straighten the furrow according to the line and wraps the parts of the image equally distributed on the left and right of the furrow (\cref{fig:Figure1}E).

\subsection{Cell Shape Segmentation in 2D and 3D}
The masked 3D raw images and the 2D projections of the apical and basal surface layers can be used for cell segmentation. We implemented a graphical user interface that allows to semi-automatically segment cells in the 2D projection images. A classical 2D watershed was initialized from h-minima derived seeds \cite{Meyer1994, Soille13}. Parameters can be interactively adjusted and the GUI allows to add and remove seeds for further improving the segmentation results. In addition  to the 2D projections, the obtained masked raw images can also be used for more extensive cell shape analyses in 3D+t and can directly be used as input for fully-automatic 3D segmentation methods if image quality suffices \cite{Mosaliganti12, Khan14, Stegmaier16g, Stegmaier18}. In order to measure cell shape properties in image data with limited image quality where fully automatic methods still fail to provide error-free segmentations, we employed the interactive graphical user interface SEGMENT3D by Spina \etal \cite{Spina18} that allows the user to load small 3D crops, to manually draw scribbles within the cells of interest and to interactively correct remaining errors. A marker-based 3D watershed algorithm then uses the manual annotations to automatically compute the pixel-accurate segmentations. The semi-automatically generated tightly masked raw images (\cref{sec:MaskGeneration}) could be perfectly used as input to the tool. The introduced safety margins at the embryo to background interfaces are particularly helpful for the watershed-based analysis and help preventing interior regions from flooding across the apical or basal membranes and to obtain a single connected component for the background.

\subsection{Automatic Tracking and Measurement}

We implemented an automatic tracking and quantification tool in MATLAB to postprocess 2D segmentations or manually segmented consecutive 3D image snippets. For the 2D projections, we incorporated the nearest neighbor-based tracking algorithm of the SciXMiner toolbox \cite{Mikut17}. For each tracked cell we extract quantitative features like area, eccentricity and speed. For the 3D data, the tracking was performed by first extracting all connected components from the manually segmented image snippets separately for each of the frames and then propagating the IDs of cells in the first frame to the successive frames by matching objects with the largest spatial overlap. Only those objects that were successfully and unambiguously tracked over the entire time span were used for the subsequent quantifications. Thus, objects at the image boundaries that potentially leave the region of interest during the analyzed time span, were discarded. We then measured the volume of each cell as the number of voxels, scaled by the physical extents of each voxel. Moreover, the apical surface of each cell was extracted by shrinking the outer surface using morphological dilation operation with a spherical structuring element with a radius of 2 pixels and by subtracting the shrinked binary image from the initial binary image. The binary mask of the outer shell was then multiplied with the original label image. This procedure yielded the apical surface of the current snippet. To measure the areas, we separately projected the apical surface voxels of each cell to a 2D plane spanned by the first two principal component vectors of the spatial positions of the apical surface voxels. Combined with the single-cell tracking, this allowed to quantify the temporal changes of the cell volume and apical areas, respectively, separately for each of the cells.

\section{Analyzing Drosophila Gastrulation}
\begin{figure}[htb!]
\includegraphics[width=\columnwidth]{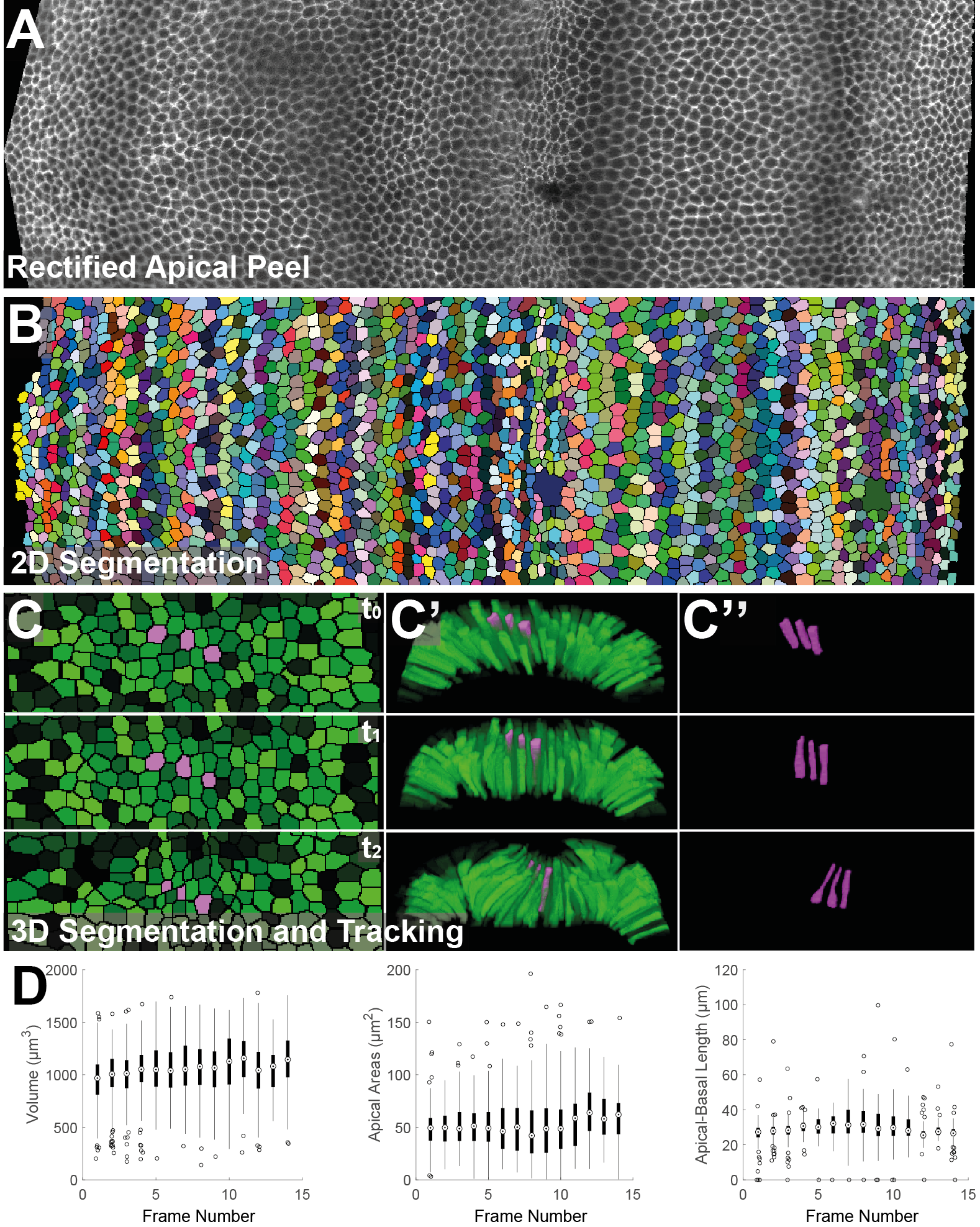}
\vspace{-0.7cm}
\caption{Application of the presented modules. (A) Apical peel after applying the rectification tool, (B) watershed-based 2D segmentation including the possibility to manually correct seed points. (C) Fully-automatic 3D tracking results using manually masked 3D images and the algorithm described in \cite{Stegmaier18}. Segmentation results are visualized as 2D slice (C), 3D volume rendering (C') and a 3D volume rendering of a subset of the cells highlighted in magenta (C'') for three selected frames. (D) Quantification of volume, apical area and apical-basal length extracted from the 3D+t data.}
\label{fig:Figure2}
\end{figure}

We applied the presented tools to 3D+t light-sheet microscopy data of a \emph{Drosophila} embryo comprising $16$ frames imaged at $40$\,s intervals. For observing cell membranes, GAP43::Venus (Bloomington stock no. 30896) embryos were imaged using a Luxendo MuVi-SPIM from two opposing directions simultaneously and successively from two directions with 90 degree apart. Illumination was done using Nikon 10/0.3W objective lenses and detection with Nikon 20/1.0W objective lenses. 1.5X magnification was used to obtain an effective voxel size of $0.19 \operatorname{\mu m} \times 0.19 \operatorname{\mu m} \times 0.5 \operatorname{\mu m}$. Images were registered, transformed and fused with the Luxendo Image Processor \cite{Luxendo19}. 3D volumetric segmentation and quantification (\cref{fig:Figure2}C, D) were performed on GAP43::Venus embryos, which were imaged using 2-photon illumination with a femtosecond-pulsed
infrared laser (Chameleon Compact OPO Family, Coherent) tuned at
950 nm emission wavelength and coupled with a Zeiss LSM 780 confocal
microscope with C-Apochromat
63X 1.4NA objective.

The apical peel depicted in \cref{fig:Figure2}A was extracted from a semi-automatically masked image captured at 3\,min after the onset of gastrulation and exhibits a clearly non-convex surface (masking procedure takes about 2\,min per time point). Due to the 3D boundary tracing algorithm, areas in the 2D projection will be faithfully preserved. In \cref{fig:Figure2}B, we applied a seeded watershed algorithm to the 2D projection image. Seeds were semi-automatically identified with an interactively adjustable h-minimum transform and the possibility to manually add or remove detections, respectively ($3-5$\,min of correction per slice). After all frames are segmented properly, the 2D segments can be tracked over time, \eg, to qualitatively analyze the spatial rearrangement of the cells or to quantify temporal changes of shape features on the single cell level (data not shown). Moreover, the semi-automatically masked raw images can be further processed by manual \cite{Spina18} or automatic 3D segmentation approaches \cite{Stegmaier18} and the proposed tracking module can be used to establish temporal correspondences between the cells via a largest spatial overlap of successive frames (\cref{fig:Figure2}C). The tracked 3D cells can be used to analyze temporal changes of measures like cell volume, apical area and the apical-basal length (\cref{fig:Figure2}D). 

\section{Conclusion}
In this contribution we presented a set of tools that can be used for a semi-automatic analysis of 3D+t microscopy images of early \emph{Drosophila} embryos. This includes semi-automatic masking procedures, automatic projection and rectification of 3D surfaces to 2D, as well as methods for manual, semi- and fully-automatic segmentation and tracking. Particularly, the presented tools allow analyzing even non-convex embryo shapes with the possibility to manually intervene all processing steps in case of poorly resolved image areas. All described algorithms are open-source and were implemented in MATLAB (Apache License 2.0)\footnote{Source code including examples and instructions for all processing steps is available at \url{github.com/stegmaierj/CellShapeAnalysis/} .}.


 One limitation is that steps along the anteroposterior axis are assumed to be identical, \ie, in the case of strong curvature along the AP axis, we underestimate the areas of cells in the 2D projection (which is negligible for the core region of the \emph{Drosophila}). Moreover, the semi-automatic delineation of the apical and the basal surfaces involves several manual steps and is a bottleneck when analyzing long time series. To make the proposed approach also applicable to large-scale time-resolved image data sets, a logical next step will be the automation of the foreground identification, \eg, by predicting the interfaces between embryo and surrounding medium with convolutional neural networks \cite{Cicek16, Milletari16}.

\clearpage
\bibliographystyle{IEEEbib}
\bibliography{Bibliography}

\end{document}